\documentclass[pra,twocolumn,showpacs,floatfix,amsmath,amsfonts]{revtex4}
\usepackage{amsmath}
\usepackage{epsfig}
\usepackage{amssymb}
\begin{document}

\title{Symbiotic Solitons in Heteronuclear Multicomponent Bose-Einstein condensates}

\author{V\'{\i}ctor M. P\'erez-Garc\'{\i}a}
\affiliation{Departamento de Matem\'aticas, Escuela T\'ecnica
Superior de Ingenieros Industriales, \\
Universidad de Castilla-La Mancha 13071 Ciudad Real, Spain}

\author{Juan Belmonte Beitia}
\affiliation{Departamento de Matem\'aticas, Escuela T\'ecnica
Superior de Ingenieros Industriales, \\
Universidad de Castilla-La Mancha 13071 Ciudad Real, Spain}

\begin{abstract}
We show that bright solitons exist in quasi-one dimensional heteronuclear multicomponent Bose-Einstein condensates with repulsive self-interaction and attractive inter-species interaction. They are remarkably robust to perturbations of initial data and collisions and can be generated by the mechanism of modulational instability. Some possibilities for control and the behavior of the system in three dimensions are also discussed.
\end{abstract}

 \maketitle

\section{Introduction} 

Symbiosis is an assemblage of distinct organisms living together. 
Although the original definition of symbiosis by De Bary \cite{1879} did 
not include a judgment on whether the partners benefit or harm each other, 
currently, most people use the term symbiosis to describe interactions 
 from which both partners benefit.  

In Physics,  waves in dispersive linear media tend to expand 
due to the different velocities at which the wave components propagate. 
This is not the case in many nonlinear media, in which certain 
  wavepackets,  called \emph{solitons} are able to propagate undistorted 
  due to the balance between dispersion and nonlinearity \cite{Scott}. 
  
Stable solitons of different subsystems are sometimes able to ``live together" 
 and form stable complexes called vector solitons
 as it happens with Manakov optical solitons \cite{Manakov,Kivshar} or stabilized vector solitons \cite{PGVector}. 
 In some cases,  a (large) robust soliton can be used to stabilize a (small) weakly unstable wave  \cite{saturable}.  
  
Multicomponent solitary waves also appear in Bose-Einstein condensates (BECs). In fact, multicomponent BECs support nonlinear waves which do not exist in single component BECs such as  domain wall solitons \cite{Cohen,Kasamatsu}, dark-bright solitons \cite{Anglin}, etc. Most of the previous analyses correspond to homonuclear multicomponent condensates for which the atom-atom interactions are repulsive. However, heteronuclear condensates offer a wider range of possibilities, the main one being the possibility of having a negative inter-species scattering length. This possibility has been theoretically explored in the context of Feschbach resonance management \cite{Simoni} and realized experimentally for boson-fermion mixtures \cite{KRb,NaLi}.

In this paper we study the existence and properties of bright solitons  in heteronuclear two-component BECs with scattering lengths $a_{11}, a_{22} > 0 $ and $a_{12} <0$. 
We would like to stress the fact that these coefficient combinations do not arise in other systems where similar model equations are used. For instance in nonlinear optics, where the nonlinear Schr\"odinger equations used to describe the propagation of laser beams in nonlinear media are similar to the mean field equations used to describe Bose-Einstein condensates, the nonlinear coefficients are allways of the same sign. The closest analogy could happen in the so-called 
 QPM ( quasi-phase-matched ) quadratically nonlinear media, where an \emph{effective} cubic nonlinearity could  be ``engineered" which could have 
 similar properties  but we do not know of any systematic studies of those systems.

Our analysis will show novel features with respect to those already found in single species BECs \cite{bright}. For instance, even when solitons do not exist for each of the species, the coupling leads to robust vector solitons. Since the mutual cooperation between these structures is essential for their existence we will refer to these solitons hereafter as \emph{symbiotic solitons}. We also show how they appear by modulational instability and study some features of their collisions. We also comment on the possibility of obtaining these structures in multidimensional configurations.

\section{The model and its basic properties}
In this paper we will study two-component BECs in the limit of strong transverse confinement ruled by \cite{Perez-Garcia}
\begin{subequations} \label{model1}
\begin{eqnarray}
i \frac{\partial u_1}{\partial t} & = & -\frac{1}{2} \frac{\partial^2 u_1}{\partial x^2} + \left(g_{11}|u_1|^2+g_{12}|u_2|^2\right)u_1, \\
i \frac{\partial u_2}{\partial t} & = & -\frac{\alpha}{2} \frac{\partial^2 u_2}{\partial x^2} + \left(g_{21}|u_1|^2+g_{22}|u_2|^2\right)u_2,
\end{eqnarray}
\end{subequations}
 where $x$ is the adimensional longitudinal spatial variable measured in units of  $a_0 = \sqrt{\hbar/m_1\omega_{\perp}}$, 
$t$ is the time measured in terms of  $1/\omega_{\perp}$, and $u_j({\bf x},t) \equiv u_j({\bf
r},\tau)\sqrt{a_0^3}$. The dimensional reduction leads to \cite{Perez-Garcia} $g_{ij} = 2 a_{ij}\alpha^{i+j-2}/a_0$, with $\alpha = m_1/m_2$ and $a_{ij}$ being
the $s-$wave scattering lengths. The normalization for $u_j$ is $\int
|u_j|^2 d^3x = N_j$ where $N_j$ is the number of particles
of each species.

 Let us first consider constant amplitude solutions of Eq. (\ref{model1}), which are of the form 
 \begin{subequations}
\begin{eqnarray}
\phi_j(z,t) & = & A_j e^{i\beta_jt},\\
 \beta_j & = & g_{jj}|A_j|^2+g_{j,3-j}|A_{3-j}|^2,
\end{eqnarray}
\end{subequations}
for $j=1,2$.  We will study the evolution of small perturbations of $\phi_j$ of the form
\begin{equation}
u_j(z,t) = \left(A_j + \delta A_j(z,t)\right)e^{i (\beta_j t + \delta \beta_j(z,t))}
\end{equation}
Using Eq. (\ref{model1}) and retaining the first order terms we get partial differential equations for $\delta A_1, \delta \beta_1, \delta A_2, \delta \beta_2$ which can be transformed to Fourier space to obtain
\begin{equation} 
\delta A_j(z,t) = \int_{\mathbb{R}} a_0(k) e^{ikx} e^{\Omega(k)t} dk
\end{equation}
$a_0(k)$ being the Fourier transform of the initial perturbation. Perturbations remain bounded if $\text{Re} \left[ \Omega(k) \right] \leq 0$. Some algebra leads to
\begin{equation} 
\Omega^2 = \tfrac{1}{2}\left(f_1+ f_2 \pm \sqrt{(f_1-f_2)^2+4C^2}\right)
\end{equation}
where $f_j = -(g_{jj}A^2_j+k^2/4)k^2, C^2 = A_1^2A_2^2 g^2_{12}k^4$.
The so-called modulational instability (MI) occurs when $\Omega(k)^2 >0$ for any $k$. For small wavenumbers (worst situation) we get
 \begin{equation}\label{MI}
g_{12}^2 > g_{11}g_{22},
\end{equation}
which  is analogous to the miscibility criterion for two-component condensates \cite{Kasamatsu}. 
However, the physical meaning of Eq. (\ref{MI}) is very different since now this instability is a signature of the tendency to form coupled 
objects between both atomic species. 
The role of MI in the formation of soliton trains and domains in BEC has been recognized in previous papers \cite{Kasamatsu,Min1,bright}.

\section{Vector solitons}

Eqs. (\ref{model1}) have sech-type solutions 
\begin{equation}\label{solitons}
u_j(x,t) = \left(\frac{N_j}{2\omega}\right)^{1/2} \text{sech} \left(\frac{x}{\omega}\right) e^{i \lambda_j t}
\end{equation}
with $ \lambda_1 = 1/(2\omega^2), \lambda_2 = \alpha/(2\omega^2)$, and $\omega = 2/\left(-g_{11}N_1-g_{12}N_2\right),$ provided  the restriction
\begin{equation}\label{restriction}
g_{12} \left(m_1 N_2- m_2 N_1 \right) = m_2 g_{22} N_2 - m_1 g_{11} N_1,
\end{equation}
and the MI condition (\ref{MI}) are satisfied. Eq. (\ref{restriction}) implies that, given the number of particles in one component the other is fixed. 

Since the self-interaction coefficients are positive,  these solitons are supported only by the mutual attractive interaction between both components. This type of vector soliton thus differs from others described for Nonlinear Schr\"odinger equations of the form Eq. (\ref{model1}), such as the Manakov solitons \cite{Manakov}, where all the nonlinear coefficients cooperate to form the solitonic solution.
\begin{figure}
\epsfig{file=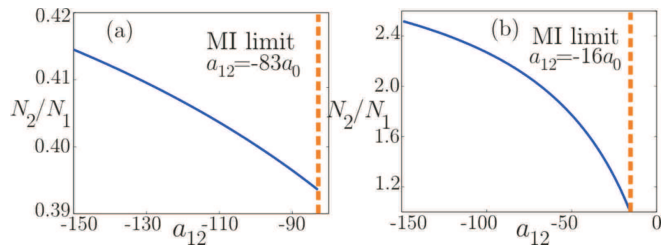,width=\columnwidth}
\caption{Dependence of the ratio $N_2/N_1$ of sech-type vector solitons on the inter-species scattering length $a_{12}$. (a) A $^{87}$Rb-$^{41}$K mixture with $a_{11} = 69a_0, a_{22} = 99a_0$, (b) An hypotetical $^7$Li-$^{23}$Na mixture with $a_{11} = 5 a_0$ and $a_{22} = 52 a_0$. \label{plota}}
\end{figure}
\begin{figure}
\epsfig{file=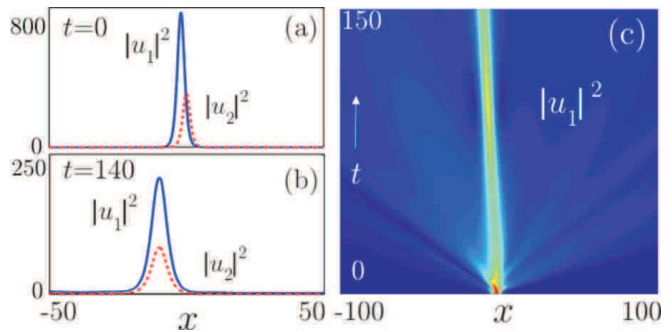,width=\columnwidth}
\caption{[Color online] Evolution of displaced soliton initial data of the form 
$u_1= (N_1/(2w))^{1/2}\text{sech}\left((x+x_0)/w\right)$, $
  u_2 = (N_2/(2w))^{1/2}\text{sech}(x/w)$ 
 for a $^{87}$Rb-$^{41}$K mixture with $N_1 = 3000, N_2 = 1189, a_{11} = 69 a_0, a_{12}=-90 a_0,  a_{22}= 99 a_0$ in a trap with $\omega_{\perp} =$ 215 Hz. \label{dual}}
\end{figure}
The MI condition (\ref{MI}) implies that the formation of these solitons has a threshold in $g_{12}$ and means that
the cross-interaction must be strong enough to be able to overcome the self-repulsion of each atomic cloud. There are no analogues to this condition in single component systems since solitons exist for any value of the self-interaction coefficient $g<0$.
To fix ideas, taking a $^{87}$Rb-$^{41}$K mixture with $a_{11} = 69a_0, a_{22} = 99a_0$ the MI condition implies that $a_{12} < - 83 a_0$ in order to obtain solitons. In Fig. \ref{plota}(a) it can be seen how the ratio $N_2/N_1$  is close to 0.4 in the range of values of $-83 a_0 >a_{12} > -150 a_0$. An hypothetical $^7$Li-$^{23}$Na mixture with $a_{11} = 5 a_0$ and $a_{22} = 52 a_0$ (in appropriate quantum states) leads to the curve in Fig. \ref{plota}(b), which shows a much larger range of variation.

\section{Soliton Stability} 
 We can use the Vakhitov-Kolokov (VK) criterion 
 to study the stability of solitons given by Eq. (\ref{solitons}). To do this, we must study the sign of $\partial \lambda_j/\partial N_j$. For soliton solutions  this can be done from the explicit form of $\lambda_j$. After some algebra we find $\lambda_1(N_1)$ and $\lambda_2(N_2)$ and obtain that
$\partial \lambda_1/\partial N_1 > 0,$ and
$\partial \lambda_2/\partial N_2  > 0$ in all their range of existence, which \emph{proves the linear stability} of the solitons for small perturbations and contradicts the naive intuition that the self-repulsion would lead to intrinsically unstable wavepackets. 

We have studied numerically the robustness of symbiotic solitons to finite amplitude perturbations. First we have perturbed both solutions with small amplitude noise and found that, in agreement with the predictions of the VK criterion, they survive after the emission of the noise in the form of radiation. Next we have applied a stronger perturbation consisting of displacing mutually their centers and observe that a soliton is formed even for relative displacements of the order of the soliton size [Fig. \ref{dual}]. Finally we have started with sech-type initial data which are not solitons and observe that after the emission of some radiation  solitons are formed.

\section{Generation of symbiotic solitons by MI} To study the generation of these solitons by MI in realistic systems we have considered a multicomponent Bose-Einstein condensate of $^{87}$Rb and $^{41}$K atoms for which the inter-species scattering length $a_{12}$ is controlled by the use of Feschbach resonances as proposed in \cite{Simoni}.  To simplify the problem here we do not consider the effect of gravity. 

\begin{figure}
\epsfig{file=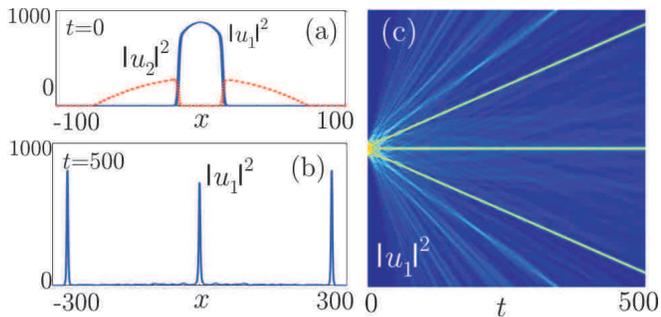,width=\columnwidth}
\caption{[Color online] Evolution of the ground state of a $^{87}$Rb-$^{41}$K mixture with $N_1 = 25000, N_2 = 20000$ after switching the interspecies scattering length from $a_{12} = 95a_0$ to $a_{12} = -90a_0$.  (a)  Initial state: $|u_1|^2$ (blue dotted line) and
$|u_2|^2$ (red solid line). (b) Profile of $|u_1|^2$ for $t=500$ showing three remaining solitons. (c) Pseudocolor plot of $|u_1|^2$ for $x\in [-500,500]$ and $t \in [0,500]$.
\label{solitrains}}
\end{figure}

We start by constructing the ground state of the system for an elongated trap typical of the LENS setup with  $\omega_{\perp} =$ 215 Hz, $\omega$ = 16.3 Hz. For these atomic species $a_{11} = 69 a_0$ and $a_{22} = 99 a_0$. We adjust the inter-species scattering length to $a_{12}=95 a_0$ during the condensation process. The ground state of this system for $N_1 = 25000, N_2 = 20000$, shown in Fig. \ref{solitrains}(a), agrees well with the theoretical predictions for these systems \cite{BT}.

\begin{figure}
\epsfig{file=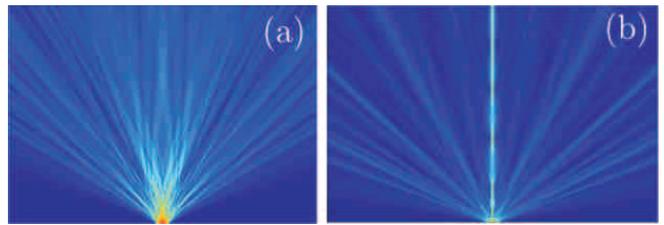,width=\columnwidth}
\caption{[Color online] Pseudocolor plots of the evolution of the ground state of a $^{87}$Rb-$^{41}$K mixture with $N_1 = 25000, N_2 = 20000$ after switching the inter-species scattering length from $a_{12} = 95a_0$ to: (a)  $a_{12} = -70a_0$ (below the MI limit); (b) $a_{12} = -87a_0$ (slightly above the MI limit $a_* = -83a_0$.
\label{hilos}}
\end{figure}
After the condensate is formed we change instantaneously this quantity to a negative value and at the same time switch off the longitudinal trapping potential and
observe numerically the evolution of the ground state. 

First we choose $a_{12} = -90 a_0$ and observe the evolution starting from the ground state with $a_{12} = 95$. Since the inter-component repulsive force is not present now, the sharp domain wall separating both species (see Fig. \ref{solitrains}(a)) decay  through a highly oscillatory process related to the formation of a shock wave \cite{Kuzmiak}. The final outcome is the formation of a soliton train (see Fig. \ref{solitrains}(b,c))
of which three solitons of about 20 $\mu$m size and each with about 3000 rubidium and 1200 potassium atoms remain in our simulation domain after 500 adimensional time units [Fig. \ref{solitrains}(c)]. Other smaller and wider solitons exit our integration region traveling at a faster speed.

The final number of solitons depends on the value of $a_{12}$ choosen during the condensation process (which controls the overlapping of the species) and the number
of particles $N_1$, $N_2$ and the negative scattering length $a_{12}$ choosen to destabilize the system. For instance, choosing $a_{12} = -70 a_0$, which is below the theoretical limit for MI the evolution of the wavepacket  is purely dispersive [see Fig. \ref{hilos}(a)]. Choosing $a_{12} = - 87 a_0$, above the MI limit but below the choice of Fig. \ref{solitrains} leads to the formation of a single soliton [Fig. \ref{hilos}(b)]. It seems that the larger the scattering length, the larger the number of solitons which arise after the decay of the initial configuration. The many degrees of freedom present in these system open many posibilities for controling the number and sizes of solitons by appropriately choosing the values of $a_{12}$ before and after the condensate is released and the initial number of particles $N_1, N_2$.

\begin{figure}
\epsfig{file=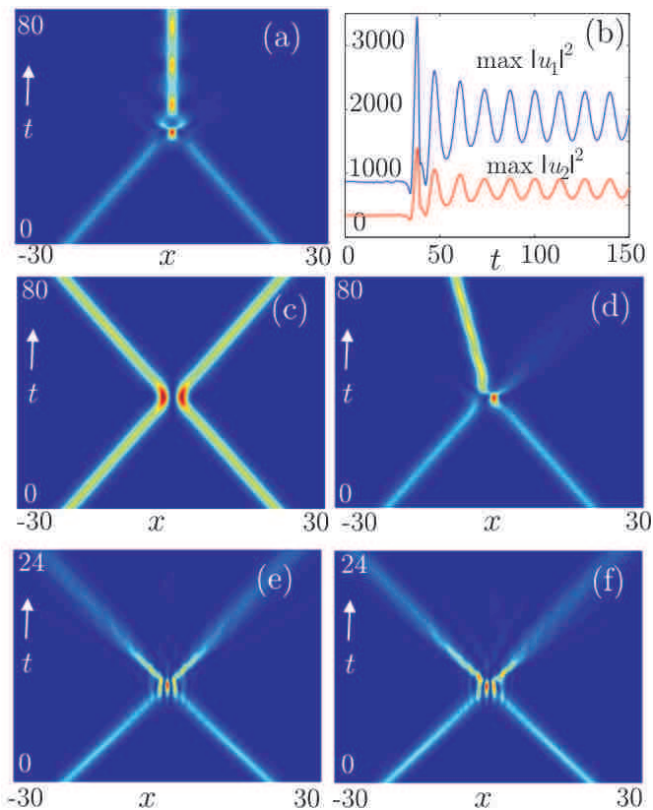,width=\columnwidth}
\caption{[Color online] Head-on collisions of symbiotic solitons with $N_1 = 3000, N_2 = 1189, w= 1.723,  a_{11}=69a_0,  a_{12}=-90a_0$ and  $a_{22}=99a_0$. (a)-(d) Slow collisions for $v=0.05$ and (a-b) $\boldsymbol{\theta} \equiv (\alpha_{1,+},\alpha_{1,-},\alpha_{2,+},\alpha_{2,-}) = (0,0,0,0)$, (c) $\boldsymbol{\theta}  = (0,\pi,0,0)$, (d) $\boldsymbol{\theta}  = (\pi/2,0,0,0)$.  Moderate speed collisions (e)  $v=0.2, \boldsymbol{\theta} = (0,0,0,0)$, (f) $v = 0.2, \boldsymbol{\theta}  = (\pi/2,0,0,0)$. \label{tortazo}}
\end{figure}

\section{Collisions of symbiotic solitons} The robustness of symbiotic solitons manifests also in their collisional behavior and their internal structure makes the interaction of these vector solitons very rich. Since each soliton is a compound object the collisions are at the same time a coherent phenomenon because of the direct overlapping of the same type of atoms and an incoherent one because of the incoherent nature of interaction between different types of atoms. A related subject of recent interest in Optics is that of partially coherent solitons \cite{partial}.

We have simulated head-on collisions of equal symbiotic solitons of opposite velocities given by 
\begin{eqnarray}
u_j & =&  \sqrt{\frac{N_j}{2w}}\text{sech}\left(\frac{x+x_0}{w}\right)e^{iv\sqrt{m_j}x+i\alpha_{j,+}} \nonumber \\
& +  & \sqrt{\frac{N_j}{2w}} \text{sech}\left(\frac{x-x_0}{w}\right)e^{-iv\sqrt{m_j}x+i\alpha_{j,-}}
\end{eqnarray}
for $j=1,2$. $ \alpha_{j,\pm}$ are the relative phases and
$N_2$ is given by Eq. (\ref{restriction}). In Fig. \ref{tortazo} we show some examples of these collisions. Slow [Fig. \ref{tortazo}(a-d)] or moderate speed collisions [Fig. \ref{tortazo}(e-f)] lead to bound solitons while for larger speeds the picture is not so clear. The specific outcome of the collision depends on the relative soliton phases with the phase difference between the larger components in the symbiotic soliton (in this case Rb) being the dominant ones. For instance  collisions with phases $\boldsymbol{\theta} \equiv (\alpha_{1,+},\alpha_{1,-},\alpha_{2,+},\alpha_{2,-}) = (0,\pi,\pi,0)$ [Fig. \ref{tortazo} (c)] and $\boldsymbol{\theta} = (0,0,\pi,0)$ (not shown) both lead to mutual repulsion 
but the outgoing speeds are different due to the different interactions between the internal components of the soliton. Collisions with higher but still moderate speeds [Fig. \ref{tortazo}(e-f)] give independent vector solitons. The outcome of the collisions with zero phases
is a bound state of two vector solitons which has internal oscillations, i.e. some sort of multicomponent higher order soliton.

\section{Prospects for Multidimensional Symbiotic Solitons} 

A very interesting question arising naturally is: do these symbiotic solitons exist in multidimensional scenarios? In principle the answer is not evident since 
the only effect acting against stabilization of multidimensional soliton structures would be collapse, but one could think that in this case collapse could be inhibited because of the 
\emph{repulsive} self-interaction, thus a deeper analysis is in order.

 The adimensional model equations in two and three dimensions take the form
\begin{equation}
i \frac{\partial u_j}{\partial t} =  \left(-\frac{1}{2m_j} \Delta   + V_j + g_{j,j}|u_j|^2 + g_{j,k}|u_{k}|^2\right)u_{j},
\end{equation}
with $j = 1,2$ and $k = 2,1$ correspondingly. 

Let us first consider this problem in two spatial dimensions. To study collapse rigorously one usually tries to compute  the exact evolution of the wavepacket widths rigorously \cite{PG99}. For the multicomponent case and $m_1 = m_2 = m$, this was studied by group-theoretical methods  by \cite{Gosh}. In our case, from the general formulae obtained by Gosh we get a sufficient condition for collapse, which is 
\begin{multline}
\mathcal{H} = \int_{\mathbb{R}^n} \left [\sum_{j=1,2} \left(|\nabla u_j|^2/(2m) +   V_j|u_j|^2 \right.\right.\\
 + \left.g_{jj}|u_j|^4/2\right) + \left.g_{12}|u_1|^2 |u_2|^2 \right]< 0.
 \end{multline}
 In principle, this is a bad result for obtaining localized structures since it means that arbitrarily close to any stationary solution
  (for which $\mathcal{H} = 0$), there would be collapsing solutions and thus stationary solutions, if they exist, would be unstable.
  As it is usual in the framework of collapse problems the situation would be even worse in three spatial dimensions with solutions of
  arbitrary small number of particles undergoing collapse provided they are initially sufficiently localized.
  
  This means that in principle symbiotic solitons could only be obtained in quasi-1D geometries because of the transverse stabilization
  effect provided by the trap in a similar way as ordinary bright solitons do.

\section{Conclusions and extensions}

 In this paper we have studied  vector solitons in heteronuclear two-component BECs which are supported by their attractive mutual interaction. These symbiotic solitons are linearly stable and remarkably robust and can be generated through modulational instability phenomenon with many possibilities for control. Collisions of these vector solitons show their robustness and open different ways for their manipulation and the design of novel quantum states such as breather-like states. We have also considered multidimensional configurations and shown that collapse may avoid the formation of 
fully multidimensional symbiotic solitons.

We think that the conceptual  ideas behind our work can also be used  to understand boson-fermion mixtures. For instance, $a_{12}$ is known to be negative and large for quantum degenerate mixtures of  $^{87}$Rb and $^{40}$K \cite{FB2}. In those systems  numerical simulations have proven the formation of localized wavepackets \cite{Karpiuk} which could share the same essential mechanisms for the formation of solitary waves. 

\acknowledgments

We acknowledge V. Vekslerchik, R. Hulet and B. Malomed for discussions.This work has been partially supported by grant BFM2003-02832
(Ministerio de Educaci\'on y Ciencia, Spain).

\end{document}